\documentclass[aps,prl,preprintnumbers,showpacs,nofootinbib]{revtex4}
\usepackage{graphicx}
\usepackage{dcolumn}
\usepackage{bm}
\usepackage{subfig}
\begin{document}
\newcommand{\newc}{\newcommand}
\newc{\ra}{\rightarrow}
\newc{\lra}{\leftrightarrow}
\newc{\beq}{\begin{equation}}
\newc{\eeq}{\end{equation}}
\newc{\barr}{\begin{eqnarray}}
\newc{\earr}{\end{eqnarray}}
\newcommand{\Od}{{\cal O}}
\newcommand{\lsim}   {\mathrel{\mathop{\kern 0pt \rlap
  {\raise.2ex\hbox{$<$}}}
  \lower.9ex\hbox{\kern-.190em $\sim$}}}
\newcommand{\gsim}   {\mathrel{\mathop{\kern 0pt \rlap
  {\raise.2ex\hbox{$>$}}}
  \lower.9ex\hbox{\kern-.190em $\sim$}}}
  \def\rpm{R_p \hspace{-0.8em}/\;\:}
\title {Prospects of detection of relic antineutrinos by resonant absorption in electron capturing nuclei.}
\author{ J. D. Vergados$^{1}$ and Yu. N. Novikov$^{2,3}$ }
\affiliation{1 University of Ioannina, Ioannina, GR 45110, Greece;
e-mail: vergados@uoi.gr\\
2 Petersburg Nuclear Physics Institute, 188300, Gatchina, Russia.\\3  St. Petersburg State University, 198504, St. Petersburg, Russia.}
\vspace{0.5cm}

\begin{abstract}
We consider the  possibility of detecting  relic anti-neutrinos by their resonant absorption  in a nucleus, which is capable of  undergoing electron capture. Ongoing and future developments in Penning Trap Mass Spectrometry and  Cryogenic Micro-Calorimetry may bring this
 possibility closer to reality. 
\end{abstract}
\maketitle
\section{Introduction}
 The detection of the relic neutrinos is of fundamental importance. These neutrinos have a very low average energy $E_{\nu}=0.00055$ eV, which corresponds to a temperature of $T=1.95~^0$K. We know that there exist three non degenerate neutrino mass eigenstates with  mass squared differences known from neutrino oscillation experiments. The resulting  mass differences are less than $\approx 0.05$ eV, so that they cannot be resolved in the planned nuclear experiments. The mass scale for these neutrinos is not known, but we will take to be $m_{\nu}=1$eV. So all neutrinos are non relativistic.\\The possibility of detecting these neutrinos in their scattering with a $^{3}$H in the KATRIN experiment has been proposed \cite{Cocco08,relic13,LiXing10}. In this note we will consider the possibility of detecting relic antineutrinos in an exothermic reaction, induced electron capture, e.g.: 
\beq
{\bar \nu}_e+(A,Z,Z_e)\rightarrow (A,Z-1,Z_e-1)^*, 
\label{Eq:sterilenu}
\eeq
where Z is, as usual,  the charge of the nucleus and $Z_e$ is the number of electrons in the atom. In studying this process, which will be called antineutrino absorption, we will try to adopt a more realistic approach than previously \cite{Cocco09,LusVig10,LiXing11}, exploiting recent experimental developments in Penning Trap Mass Spectrometry (PT-MS) and  cryogenic micro-calorimetry (MC).

\section{The formalism}
\begin{figure}[!ht]
 \begin{center}
\includegraphics[scale=0.5]{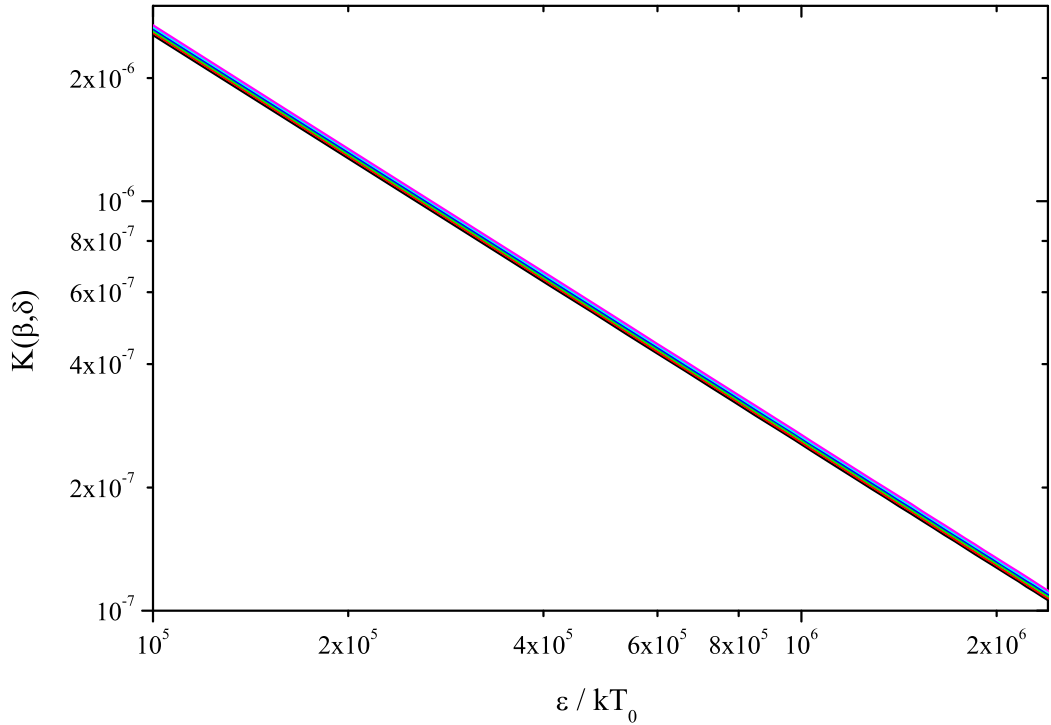}\\
 \caption{ The function $K(\beta,\delta)$ is exhibited as function of $\beta$ for the specific values of $\delta= (0., 0.02, 0.04, 0.06, 0.08, 0.1)$. The dependence on the parameter $\delta$ is not visible. Note also that the units of $\beta=\frac{\epsilon}{ k T_0}$ are in $10^{6}$.}
 \label{Fig:nuabs}
  \end{center}
  \end{figure}

\begin{figure}[!ht]
 \begin{center}
\includegraphics[scale=0.5]{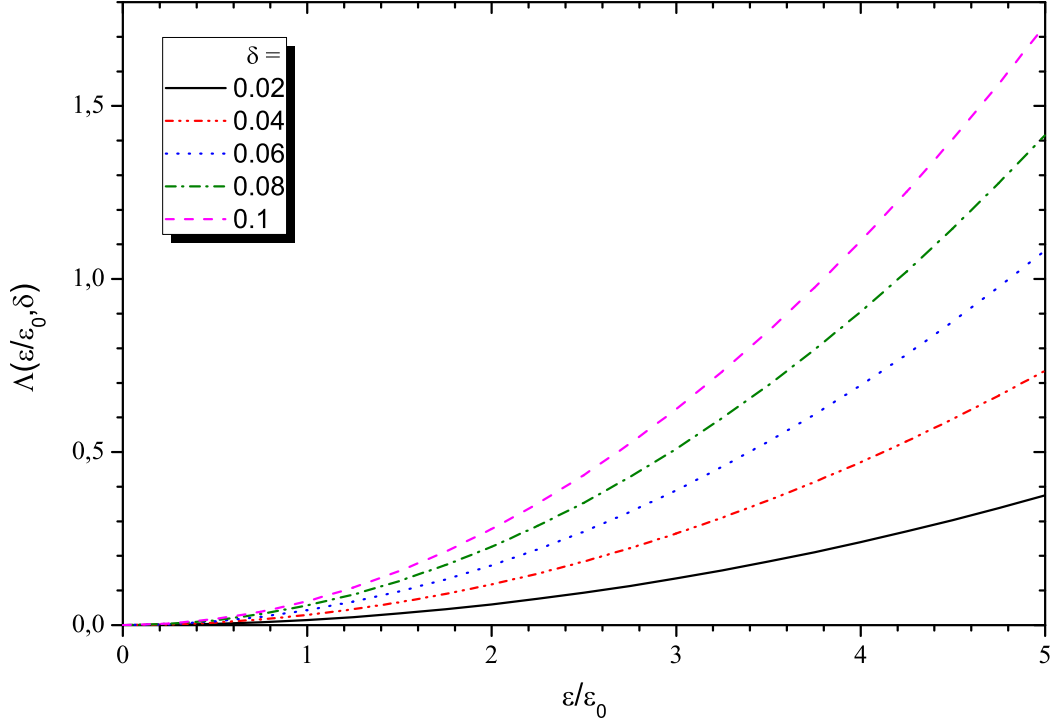}
 \caption{ The function $\Lambda(\epsilon/\epsilon_0,\delta)$ is exhibited as function of  $\epsilon/\epsilon_0$ for $\epsilon_0=0.1$ keV and the values of $\delta=(0,0.02,0.04,0.06,0.08,0.1)$ ($\delta$ is increasing upwards).}
 \label{Fig:ecapture}
  \end{center}
  \end{figure}

Let us measure all energies from the ground state of the final nucleus and assume that $\Delta$ is the mass difference of the two neutral atoms. Let us consider a transition to the final state with energy $E_x$. The cross section for a neutrino\footnote{Here as well as in the following  we may write neutrino, but it is understood that we mean antineutrino} of given velocity $\upsilon$ and kinetic energy $E_{\nu}$ is given by:
\beq
\sigma(E_{\nu})=2 \pi \frac{1}{\upsilon} |ME(E_x)|^2_{\mbox{\tiny nuc}} \prec\phi_e\succ^2\left (\frac{G_F}{\sqrt{2}} \right )^2 \delta(E_{\nu}+m_{\nu}+\Delta-E_{x}-b),
\eeq
 where $b$ the binding energy of the electron\footnote{We have neglected effects arising from  neutrino mixing, since they are not going to have a significant effect on our results.}. Clearly $E_x+b\geq\Delta+m_{\nu}$. 
The number of events per unit time per nucleus is:
\beq
{\cal N}_{\nu}(E_{\nu})=\Phi_{\nu}\sigma(E_{\nu})= n_{\nu}(r_0)\,2 \pi  |ME(E_x)|^2_{\mbox{\tiny nuc}} \prec\phi_e\succ^2\left (\frac{G_F}{\sqrt{2}} \right )^2 \delta(E_{\nu}+m_{\nu}+\Delta-E_x-b),
\eeq
with $n_{\nu}(r_0)$ the density of neutrinos in our vicinity. These neutrinos are expected to  be non relativistic gravitationally bound. This density is expected to be much higher \cite{LasVogVob08},\cite{nuden13} than the  relic neutrino density $\prec n_{\nu}\succ$. Let us suppose further that the distribution of these neutrinos is $f(E_{\nu})$. Then we can write ${\cal N}\equiv{\cal N}(\bar{\nu}-\mbox{absorption})$ as follows:
\beq
{\cal N}=\int {\cal N}_{\nu}(E_{\nu})f(E_{\nu})d E_{\nu}=\xi \prec n_{\nu}\succ\,2 \pi  |ME(E_x)|^2_{\mbox{\tiny nuc}}\prec\phi_e\succ^2 \left (\frac{G_F}{\sqrt{2}} \right )^2 f(E_x+b-m_{\nu}-\Delta),
\eeq
with $\xi=n_{\nu}(r_0)/\prec n_{\nu}\succ$ the enhancement of the neutrino density due to gravity. Since the neutrinos are bound their velocity distribution is expected  to be a Maxwell-Boltzmann distribution:
\beq
f(E)=\frac{1}{kT_0}e^{-\frac{E}{kT_0}}
\eeq
\beq
f(\upsilon)=\frac{1}{(\sqrt{\pi} \upsilon_0)^3} e^{-\left( \upsilon/\upsilon_0\right)^2},
\eeq
 with a characteristic velocity
  $$\upsilon^2_0=\frac{2 k T_0}{m_{\nu}}\approx 3 \times 10^{-4}\mbox{c}^2$$
  The numerical value was obtained using an average kinetic neutrino energy of $\epsilon_{\nu}=5.0 \times 10^{-4}$eV and a neutrino mass $m_{\nu}\approx$1eV.
i.e. higher than the corresponding value of $ k T_0=1.5 \times 10^{-4}\mbox{ eV}$ the relic background. 
   \\Thus we get for neutrino capture:
\beq
{\cal N}=2 \pi \xi |ME(E_x)|^2_{\mbox{\tiny nuc}} \left (\frac{G_F}{\sqrt{2}} \right )^2\prec\phi_e\succ^2 \frac{1}{k T_0} e^{-\frac{E_x+b-m_{\nu}-\Delta}{k T_0}}\prec n_{\nu}\succ.
\label{Eq:nuabsrate}
\eeq
This must be compared to the electron capture rate to the  final state with energy $E^{\prime}_x$:
\beq
{\cal N}(e- \mbox{capture})=\frac{1}{(2 \pi )^2} |ME(E^{\prime}_x)|^2_{\mbox{\tiny nuc}} \prec\phi_e\succ^2\left (\frac{G_F}{\sqrt{2}} \right )^2 (\Delta-E^{\prime}_x- b)^2.
\label{Eq:ecapture}
\eeq
Thus the ratio of antineutrino absorption to standard electron capture rates becomes:
\beq
\frac{\lambda_a}{\lambda_c}=\frac{{\cal N}}{{\cal N}(e- \mbox{capture})}=(2 \pi)^3\frac{|ME(E_x)|^2_{\mbox{\tiny nuc}}}{|ME(E^{\prime}_x)|^2_{\mbox{\tiny nuc}}} \xi \frac{\prec n_{\nu}\succ}{(\Delta-E^{\prime}_x- b)^2 k T_0}e^{-\frac{E_x+b-m_{\nu}-\Delta}{k T_0}}.
\eeq
We purposely chose $E_x^{\prime}\ne E_{x}$ since for the usual electron capture the final nuclear energy must be smaller than $\Delta-b$, but as we have for antineutrino absorption it must be greater than $\Delta-b+m_{\nu}$.\\
The phase space advantage of the antineutrino absorption is lost due to the kinematics and, especially, the neutrino energy distribution. One may, therefore, seek some special conditions:
\begin{itemize}
\item $E_x+b-m_{\nu}-\Delta=\epsilon,\, \epsilon \approx kT_0$. Then
\beq
\frac{{\cal N}}{{\cal N}(e- \mbox{capture})}=(2 \pi)^3\frac{|ME(E_x)|^2_{\mbox{\tiny nuc}}}{|ME(E^{\prime}_x)|^2_{\mbox{\tiny nuc}}} \xi \frac{\prec n_{\nu}\succ}{(\Delta-E^{\prime}_x- b)^2 k T_0}e^{-\frac{\epsilon}{k T_0}}.
\eeq
For $\epsilon<< k T_0$, we get
\beq
\frac{{\cal N}}{{\cal N}(e- \mbox{capture})}=(2 \pi)^3\frac{|ME(E_x)|^2_{\mbox{\tiny nuc}}}{|ME(E^{\prime}_x)|^2_{\mbox{\tiny nuc}}} \xi \frac{\prec n_{\nu}\succ}{(\Delta-E^{\prime}_x- b)^2 k T_0}.
\eeq
\item The antineutrino absorption proceeds  as above.\\ The electron capture, however, now takes from a different orbit with binding energy ${\tilde b}$. The
\beq
\frac{{\cal N}}{{\cal N}(e- \mbox{capture})}=(2 \pi)^3 \xi \frac{|ME(\Delta)|^2_{\mbox{\tiny nuc}}}{|ME( E_x^{\prime})|^2_{\mbox{\tiny nuc}}}\frac{\prec\phi_e\succ}{\prec\tilde{\phi}_e\succ}\frac{{\prec n_{\nu}\succ}}{(\Delta-E^{\prime}_x- {\tilde b})^2 k T_0}e^{-\frac{\epsilon}{k T_0}}.
\eeq

To-day the neutrinos are non relativistic, due to their finite mass, with an average energy of about $5.0\times 10^{-4}$eV (the rotational velocity in the galaxy cluster must be higher than the rotational velocity in our neighborhood in the galaxy).
Using $\prec n_{\nu}\succ=56$ cm$^{-3}$, $\Delta-E^{\prime}_x-\tilde{b}=100$ keV, $\epsilon \approx k T_0\approx 10^{-3}$eV we find:
$$(2 \pi)^3 \frac{\prec n_{\nu}\succ}{(\Delta-E^{\prime}_x-\tilde{b})^2 k T_0}\frac{1}{e}\approx 0.4 \times10^{-11}.$$
Thus, assuming nuclear matrix elements of the same order, in the first case discussed above we find:
\beq
\frac{{\cal N}}{{\cal N}(e- \mbox{capture})}=0.4 \times10^{-11} \xi .
\eeq
If however the neutrino density in our vicinity is enhanced, essentially considering gravitational clustering as a byproduct
of the clustering of dark matter  \cite{RungWang05} or assuming that the neutrino overdensity is analogous to
the baryon overdensity in galaxy clusters \cite{LasVogVob08}, one can have $\xi$ as high as $\xi=10^{6}$. Thus we get a substantial gain:
\beq
\frac{{\cal N}}{{\cal N}(e- \mbox{capture})}=4 \times10^{-6}.
\eeq 
This, however, relies on a fine tuning of the parameter $\epsilon$, hard to meet in practice.
\item{The presence of a resonance.}
Let us suppose that there is a resonance in the final nucleus at an energy $\epsilon$ above the value $\Delta-b+m_{\bar{\nu}}$ with a width $\Gamma=\epsilon(1+\delta), \delta<<1$ given by:
\beq
f(E_x)=\frac{2}{\pi}\frac{\Gamma}{\left (E_x-(\Delta-b+\epsilon+m_{\bar{\nu}})\right )^2+(\Gamma/2)^2}.
\eeq
$\delta$ is introduced so that through this resonance the ordinary electron capture may also proceed. Then, after integrating from $E_x=\Delta-b+m_{\bar{\nu}}$ to  $E_x=\Delta-b+m_{\bar{\nu}}+\Gamma$, Eq. (\ref{Eq:nuabsrate}) becomes
\beq
{\cal N}=2 \pi \xi |ME(E_x)|^2_{\mbox{\tiny nuc}} \left (\frac{G_F}{\sqrt{2}} \right )^2\prec\phi_e\succ^2 \prec n_{\nu}\succ\frac{1}{k T_0} K(\beta,\delta),
\label{Eq:nuabsres}
\eeq
with
\barr
K(\beta,\delta)&=&-\frac{1}{\pi}i e^{-\frac{1}{2} i (\delta +1) \beta -\beta }
   \left(-E_1\left(\left(1-\frac{i}{2}\right) (\delta +1) \beta
   \right)\right . \nonumber\\
  && +e^{i (\delta +1) \beta }
   \left(E_1\left(\left(1+\frac{i}{2}\right) (\delta +1) \beta
   \right)-E_1\left(\frac{1}{2} i (\delta +(1+2 i)) \beta
   \right)\right) \nonumber\\
   && \left . +E_1\left(\frac{1}{2} (\delta +1)
   \left(-i-\frac{2}{\delta +1}\right) \beta \right)\right),
\earr
where 
$$E_n(z)=\int_1^{\infty}\frac{e^{-z t}}{t^n}dt, \quad \beta=\frac{\epsilon}{ k T_0}$$
The behavior of this function is exhibited in Fig. \ref{Fig:nuabs}.\\
We should note for a resonance in this region ordinary electron capture also becomes possible. Then Eq. \ref{Eq:ecapture} becomes:
\barr
{\cal N}(e- \mbox{capture})&=&\frac{1}{(2 \pi )^2} |ME(E_x)|^2_{\mbox{\tiny nuc}} \prec\phi_e\succ^2\left (\frac{G_F}{\sqrt{2}} \right )^2 \epsilon_0^2 \Lambda(\epsilon/\epsilon_0,\delta)\nonumber,\\
\Lambda(\epsilon/\epsilon_0,\delta)&=&\frac{4}{\pi }\left (\frac{\epsilon}{\epsilon_0}\right )^2 \left(\frac{\delta }{\delta +1}+\tan
   ^{-1}\left(\frac{2}{\delta +1}\right)-\tan ^{-1}(2)\right)\approx\frac{4}{\pi}\frac{\epsilon^2}{\epsilon^2_0}\frac{3}{5}\delta \left (1-\frac{92}{25}\delta \right )\nonumber\\.
\label{Eq:ecaptres}
\earr
Thus we get:
\beq
\frac{{\cal N}}{{\cal N}(e- \mbox{capture})}=(2 \pi)^3 \xi \frac{\prec n_{\nu}\succ}{(\epsilon_0)^2 k T_0} \frac{K(\beta,\delta,)}{\Lambda(\epsilon/\epsilon_0,\delta)}.
\label{Eq:abscapt}
\eeq
“The behaviour of function $\Lambda(\epsilon/\epsilon_0,\delta)$ is exhibited in Fig.  \ref{Fig:ecapture}.
\end{itemize}

  \begin{figure}[!ht]
 \begin{center}
\includegraphics[scale=0.5]{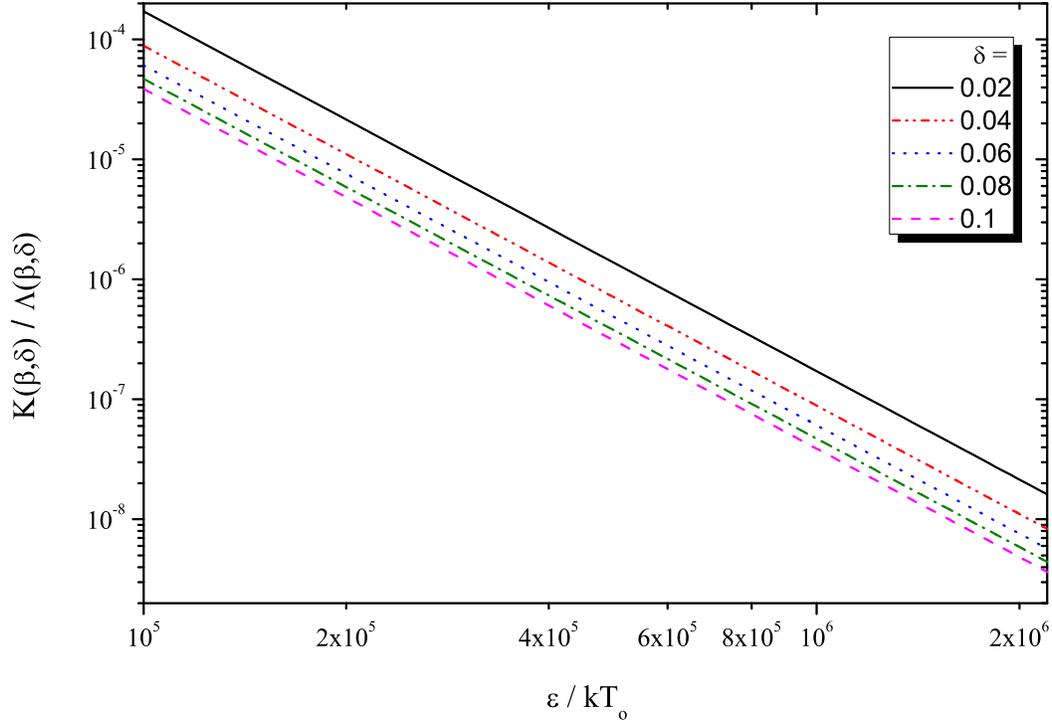}
 \caption{ The function $K(\beta,\delta)/\Lambda(\beta,\delta)$ is exhibited as function of $\beta$ for the specific values of $\delta= ( 0.02, 0.04, 0.06, 0.08, 0.1)$. Note that $\epsilon/\epsilon_0=10^{-5}\beta$ and  that the units of $\beta=\frac{\epsilon}{ k T_0}$ are in $10^{6}$.}
 \label{Fig:nuabsec}
  \end{center}
  \end{figure}
	We conclude this section by noting that the average energy available for de-excitation after ordinary electron capture  is $\Delta-b-\tilde{\epsilon},\, \tilde{\epsilon}=\epsilon\sqrt{\frac{4}{\pi} \frac{3}{5}\delta \left (1-\frac{23}{25}\delta\right )} $, i.e.  smaller than the mean energy $ (\Delta-b+\epsilon+m_{\bar{\nu}})$ resulting from antineutrino absorption.
 \section{Some results}
  For $\prec n_{\nu}\succ=56\mbox{cm}^{-3}$  and $kT_0=10^{-3}$eV we find
 $$  \frac{\prec n_{\nu}\succ}{(\epsilon_0)^2 k T_0}= 1.6 \times 10^{-13}.$$
 Let us imagine that $\epsilon=0.4$ keV.
 From Fig. \ref{Fig:nuabsec} we find $$\frac{K(\beta,\delta)}{\Lambda(\beta,\delta)}=(2.72,1.42,0.98,0.76,0.64)\times 10^{-6}, $$ for the values of $\delta= ( 0.02, 0.04, 0.06, 0.08, 0.1)$ respectively.   Thus for these values of $\delta$  Eq. (\ref{Eq:abscapt}) yields:
 \begin{itemize}
 \item $\epsilon=0.4$ keV
 $$
 \frac{{\cal N}}{{\cal N}(e- \mbox{capture})}=4.0 \times 10^{-17} \times(2.72,1.42,0.98,0.76,0.64)(\mbox{ for }\xi=1),$$
 $$ \frac{{\cal N}}{{\cal N}(e- \mbox{capture})}= 4.0 \times 10^{-11} \times(2.72,1.42,0.98,0.76,0.64)(\mbox{ for }\xi=10^6).
 $$
 \item $\epsilon=0.1$ keV.
  $$\frac{K(\beta,\delta)}{\Lambda(\beta,\delta)}=(1.74,0.90,0.63,0.49,0.41)\times 10^{-4} .$$
  Thus
  $$
 \frac{{\cal N}}{{\cal N}(e- \mbox{capture})}=4.0 \times 10^{-15} \times(2.72,1.42,0.98,0.76,0.64)(\mbox{ for }\xi=1),$$
 $$ \frac{{\cal N}}{{\cal N}(e- \mbox{capture})}= 4.0 \times 10^{-9} \times(2.72,1.42,0.98,0.76,0.64)(\mbox{ for }\xi=10^6).
 $$
 \item $\epsilon=50$ eV\\
  $$\frac{K(\beta,\delta)}{\Lambda(\beta,\delta)}=(1.39,0.72,0.50,0.39,0.33)\times 10^{-3}. $$
  Thus
  $$
 \frac{{\cal N}}{{\cal N}(e- \mbox{capture})}=4.0 \times 10^{-14} \times(2.72,1.42,0.98,0.76,0.64)(\mbox{ for }\xi=1),$$
 $$ \frac{{\cal N}}{{\cal N}(e- \mbox{capture})}= 4.0 \times 10^{-8} \times(2.72,1.42,0.98,0.76,0.64)(\mbox{ for }\xi=10^6).
 $$
 \end{itemize}
 The situation for a small width begins to approach that of the sharp state discussed above.\\
 Anyway these values are quite a bit larger than the values $7.7 \times 10^{-22},5.8 \times 10^{-22}\mbox{ and }1.4 \times 10^{-23}$ obtained \cite{LusVig10} for $Q=2.3,2.5$ and 2.8 keV respectively  for the target $^{163}$Ho and much higher \cite{Cocco08} than $6.6 \times 10^{-24}$ for tritium beta decays.\\
When $\delta$ becomes zero the electron capture is forbidden. In this case one must evaluate the nuclear matrix element and $\prec\phi_e\succ^2$ in order to obtain the rate.

   \begin{table}
\begin{center}
\begin{tabular}{  ||l | c | c | c | c|c|| }
\hline \hline
state of $^{157}$Gd&$E_f$=Energy &Population &spin-parity&$Q_{EC}-E_f-B_K$(keV)&$Q_{EC}-E_f-\prec B_L\succ$(keV)\\
& of final state & of the state&&&\\
\hline
Ground state&0&99.9\,$\%$&3/2$^{-}$&9.9-12.7&52.1-54.9\\
\hline
Excited state&54.54&$10^{-1}\,\%$&5/2$^{-}$&-&$-2.5\pm0.4$\\
\hline
\hline
\end{tabular}
\end{center} 
\caption{Decay characteristics for $^{157}$Tb (71y) $\rightarrow ^{157}$Gd(g.s.) and $^{157}$Gd (54 keV). Note that $ Q_{EC} = 60.1 -62.9$ keV \cite{Helmer04} (also planned to be measured by the Penning traps),
$b_K$ = 50.24 keV, $b_{L1}$ = 8.38 keV, $b_{L2}$ = 7.93 keV, $b_{L3}$ =7.24 keV \cite{LARKINS}.
A value of  $\prec b_L\succ = 8$ keV was assumed in this work.}
\label{table.1}
\end{table}

\section{A SPECIFIC EXAMPLE: $\bar{\nu}$-CAPTURE BY $^{157}$T$\mbox{b}$}
\begin{figure}[!ht]
 \begin{center}
\includegraphics[scale=1.0]{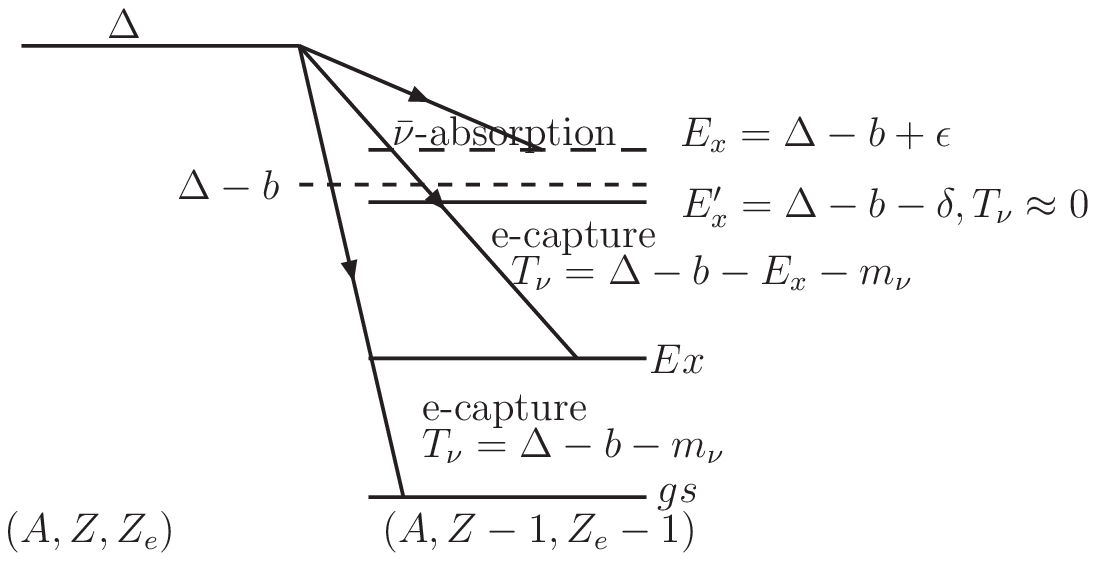}\\
 \caption{ The levels accessible to ${\bar \nu}$ absorption must be above $\Delta-b$, but for ordinary capture below. From the latter of particular interest is the level just below this value by an amount $\delta$. In this case through a nuclear  resonance with appropriate width one may have both neutrino absorption and electron capture to that state. The ratio of the corresponding rates is discussed in the text.}
 \label{Fig:levelenu}
  \end{center}
  \end{figure}

As an example we can consider the decay of $^{157}$Tb to the excited state of $^{157}$Gd with the energy 54.5 keV. The $\Delta$-value between the ground states is in the region from 60 to 63 keV \cite{Helmer04}. The level scheme appropriate for ${\bar \nu}$ absorption is given in Fif. \ref{Fig:levelenu}.
Therefore the decay energy to the excited state of 54.5 keV is equal to $Q= \Delta-54.5= 5.5 - 8.5$ keV which gives a range of -2.5 to +0.5 keV for the value of  $Q+m_{\bar{\nu}} -b_L$. Thus the resonance conditions for the antineutrino absorption by $^{157}$Tb can appear. The exact direct measurement of mass differences $\Delta$ by the Penning Trap Mass Spectrometry (PT-MS) \cite{BNW10} should show how strong this resonance is. \\
Let us assume  that the excitation energy for antineutrino capture is 0.1 keV. Then  Eq. (\ref{Eq:abscapt}) gives 
 $N/(N(\mbox{e-capture}))  = 4\times 10^{-9}$   assuming  $\xi = 10^6$.  The nuclide $^{157}$Tb can be produced in the reactor with the neutron flux of $10^{15}$ n/(cm$^2$s) on the target of the enriched 50 mg $^{156}$Dy with the production scheme
 $$^{156}\mbox{Dy+n}\rightarrow ^{157}\mbox{Dy} (T_{1/2} = \mbox{8h}) \rightarrow ^{157}\mbox{Tb}.$$
 
\begin{figure}[!ht]
 \begin{center}
\subfloat[]
{\includegraphics[scale=0.7]{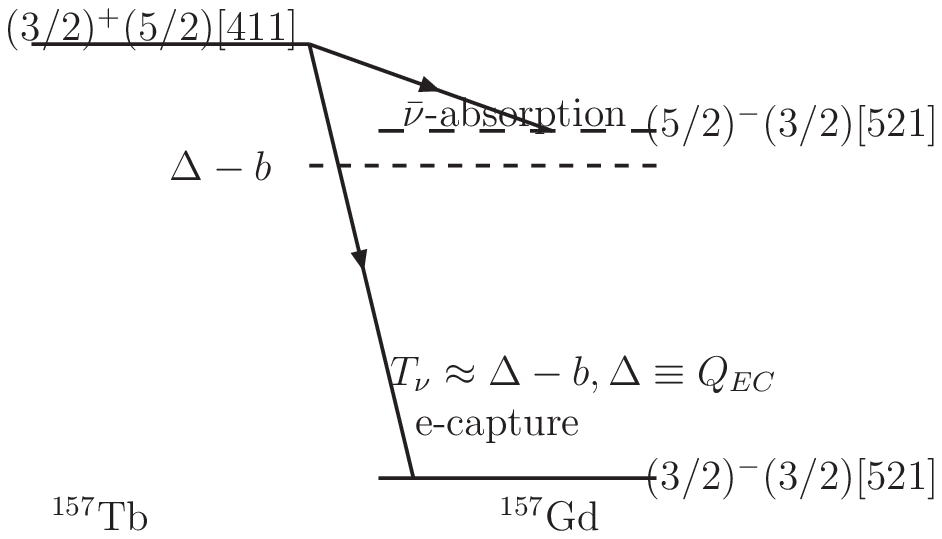}
}
\subfloat[]
{
\includegraphics[scale=0.7]{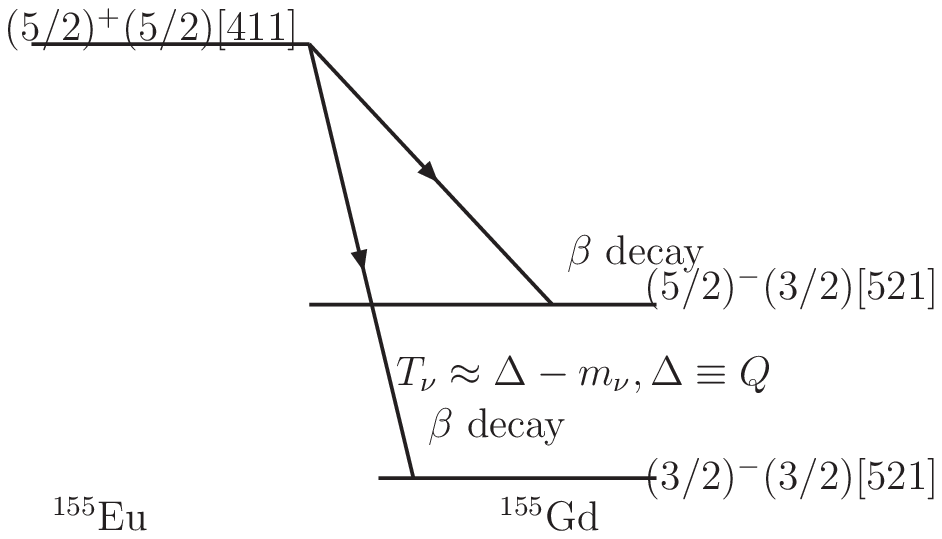}
}\\
 \caption{We show the level scheme of a nucleus relevant for ${\bar \nu}$ capture (a) and a beta decaying nucleus with similar structure, for which a calculation for the relevant nuclar matrix elements exists. In our notation  the first number is the 
 spin-parity of the state, while the band head $K$ is the second quantum number closed  in parenthesis. For the other quantum numbers see  the well known text \cite{deShFes}. We notice the similarity of the quantum numbers involved in the two systems.}
 \label{Fig:bothlevels}
  \end{center}
  \end{figure}

A few years of irradiation in the reactor will give  a number of $^{157}$Tb at the level of $10^{19}$ atoms. In order to estimate the decay rate from the ground state of $^{157}$Tb to the excited  state of 54.5 keV in $^{157}$Gd the ratios of phase-space factors, electron densities and nuclear matrix elements should be known for both transitions. Let us note that only L-capture to the excited state should be taken into account because this only  provides the resonant condition from the energy point of view. Calculation of the electron densities at the origin (atomic wave functions squared) gives \cite{Filiamin13} for the ratios of K-to L-capture a value of 7.36, which favors the ground state transition. The value of the nuclear matrix elements it is not well known. It is known that both the initial and the final nuclear systems are deformed. The gs of $^{157}$Tb is known to be of the type $(3/2)^+(3/2)[411]$ in the notation of \cite{deShFes},
 while the final states are $(3/2)^-(3/2)[521]$ and $(5/2)^-(3/2)[521]$, with the understanding that the first number is the 
 spin-parity of the state, while the band head $K$ is the second quantum number closed  in parenthesis. We notice the similarity of these quantum numbers with those of $(5/2)^+(5/2)[413]$, $(3/2)^-(3/2)[521]$ and  $(5/2)^-(3/2)[521]$ corresponding, respectively, to the ordinary beta decay of a neighboring system, namely the $^{155}\mbox{Eu}\rightarrow^{155}$Gd, for which a nuclear calculation exists \cite{Bogdan73}. The comparison is illustrated in 
Fig. \ref{Fig:bothlevels}.
 Short of a detailed nuclear calculation,  we will thus assume that the one body transition matrix elements are the same in both systems. The total nuclear ME can be expressed in terms of these via geometric factors, see e.g. Eq. (6.9), p. 409 of the well known text \cite{deShFes}. This way we find the  geometric factors favor the excited state of our system relative to the ground state by almost a factor of two. In the case of the A=155 system we see the opposite \cite{Bogdan73}, i.e. the  gs is favored by almost a factor of two. Thus we may conclude that the two nuclear matrix elements involved in our system  are about the same. 
	\\In order to estimate the branching ratio we find it convenient to express it as follows:
	\beq
	BR=\frac{\nu-\mbox{absorption}}{\mbox{e-capture to gs}}=\frac{\nu-\mbox{absorption}}{\mbox{e-capture to }E_x=\Delta-b-\tilde{ \epsilon}}\frac{\mbox{e-capture to }E_x=\Delta-b-\tilde{ \epsilon}}{\mbox{e-capture to gs}},
	\eeq
	where the first factor is given by Eq. (\ref{Eq:abscapt}). For $\epsilon=$50 eV the first factor becomes $\approx 10^{-13}\xi$.
Furthermore  we adopt the view that the branching ratio for  L-capture to the 54-keV state is  smaller than 
that dictated by the phase-space factor by a factor of 10. The obtained branching ratios are presented in table \ref{table.2}. 

\begin{table}
\begin{center}
\begin{tabular}{  ||c | c | c | c |c| c|c|| }
\hline
\hline
$m_{\nu}$(eV)&$\epsilon (keV)$&$\delta=0.02$&$\delta=0.04$&$\delta=0.06$&$\delta=0.08$&$\delta=0.10$\\
\hline
 &0.40&2.56$\times 10^{-22}$ & 2.64$\times 10^{-22}$ & 2.68$\times 10^{-22}$ & 2.73$\times 10^{-22}$ & 2.81$\times 10^{-22}$ \\
0.50&0.10& 0.96$\times 10^{-21}$ & 1.00 $\times 10^{-21}$& 1.04$\times 10^{-21}$ & 1.06$\times 10^{-21}$ & 1.09$\times 10^{-21}$ \\
 &0.05& 1.76$\times 10^{-21}$ & 1.88$\times 10^{-21}$ & 1.96$\times 10^{-21}$ & 2.08$\times 10^{-21}$ & 2.12$\times 10^{-21}$ \\
\hline
  & 0.40&2.53 $\times 10^{-22}$& 2.62$\times 10^{-22}$ & 2.67 $\times 10^{-22}$& 2.71 $\times 10^{-22}$& 2.8 $\times 10^{-22}$\\
0.75&0.10& 0.92$\times 10^{-21}$ & 0.97$\times 10^{-21}$ & 1.01$\times 10^{-21}$ & 1.04$\times 10^{-21}$ & 1.07$\times 10^{-21}$ \\
 &0.05& 1.60$\times 10^{-21}$ & 1.76$\times 10^{-21}$ & 1.86$\times 10^{-21}$ & 1.99$\times 10^{-21}$ & 2.04$\times 10^{-21}$ \\
\hline
 &0.40& 2.5 $\times 10^{-22}$ & 2.6$\times 10^{-22}$ & 2.65$\times 10^{-22}$ & 2.7$\times 10^{-22}$ & 2.79$\times 10^{-22}$ \\
1.00&0.10& 0.88$\times 10^{-21}$ & 0.94$\times 10^{-21}$ & 0.99$\times 10^{-21}$ & 1.02$\times 10^{-21}$ & 1.05$\times 10^{-21}$ \\
  &0.05&1.46$\times 10^{-21}$ & 1.65$\times 10^{-21}$ & 1.77$\times 10^{-21}$ & 1.9$\times 10^{-21}$ & 1.95$\times 10^{-21}$ \\
	\hline
	\hline
\end{tabular}
\end{center}
\caption{The branching ratio in units of $\xi$ as a function of $m_{\nu}$, $\epsilon$ and $\delta$. For the notation see text.}
\label{table.2}
\end{table}
For example for  $\epsilon=50 eV,\, \delta=0.02$ we get $\tilde{\epsilon}=6$ eV. Then $T_{\nu}= \tilde{\epsilon}-m_{\nu}=5 $ eV. We thus   get for the second factor:
   $$(1/10)\left(5/10^{4}\right )^2=2.5\times 10^{-8}$$  where $10^4$ eV  is the K-capture energy to the ground state of $^{157}$Gd. Thus with 
	$ \epsilon=$50 eV, $m_{\nu}=1 eV$ and $\xi=10^{6}$ we get a branching ratio of between  $1.46\times10^{-15}$ and $1.95\times10^{-15}$, depending on $\delta$ , i.e. between 150-200 events per year with    2.6 m-gr of source, a reasonably high value. \\
 Note, however, that this value crucially depends on the resonance condition the strength of  which can be detetermined by the measurements with the PT-MS.  
Let's assume that the method of Micro-Calorimetry (MC) \cite{Gastaldo13} could be used for identification of the relic antineutrino absorption. A measure should be the calorimetric spectrum of atomic and nuclear de-excitation after the end of the process. In this spectrum the peak corresponding to the L-capture to the 54.5-keV excited state of $^{157}$Gd should be situated at $54.5 +b_L \approx 62.5$ keV. The antineutrino absorption contribution should be observed at the tail of this peak. Other expected peaks in the spectrum are:  at $\approx8$  and 50.2 keV (respectively from the L- and K-capture to the ground state of $^{157}$Gd) and $\approx 1.5$ and 56 keV (from M-capture to the ground and to 54-keV excited state). All of these mentioned peaks are far from the peak 62.5 keV of interest,which is outermost peak in the bolometric spectrum.  This is a big advantage of the micrlocalorimetry method in the sense that the needed  definitely assigned peaks can be quite well separated with the precision of a few eV \cite{Gastaldo13}. However the disadvantage is a high pile-up background which appears when the source is too intensive. Even a multiple pixel structure of detector can require a smaller intensity of the $^{157}$Tb-source than that used in our above estimations.\\ 
Thus, in order to be able to consider the $^{157}$Tb as an appropriate candidate for testing the relic antineutrino absorption, one should, first of all, precisely measure the $\Delta$-value by means of PT-MS. This is feasible by the new generation of PT's \cite{BNW10}. Then, the development of MC,  diminishing  the pile-up background and  increasing the energy resolution, should be envisaged.    

 \section{Discussion}
 We have seen that, since the characteristic energy of the relic neutrino distribution is small, $k T_0\approx 10^{-3}$ eV, the antineutrino absorption is favored compared to electron capture, provided that  the excitation energy is small, i.e.  $\epsilon=E_x-(\Delta-b)\mbox{ a few times }k T_0$. This excited state is not available to the ordinary electron capture. Thus  the possible experimental signature is the $\gamma$ ray following the de-excitation of the nucleus and/or the observation of photons following the de-excitation of the atom. 
 \\We have found that, if there exists a resonance centered around $ (\Delta-b+\epsilon+m_{\nu})$, with a width $\epsilon(1+\delta)$, there can be a relatively large rate for antineutrino absorption. Unfortunately, however, for $\delta\ne 0$ the same mechanism feeds the ordinary electron capture to the  state just below $(\Delta-b+m_{\bar{\nu}})$ with a rate not completely suppressed. One may be able to explore experimentally the fact that, if the final state is populated by electron capture, the average energy available for de-excitation is $\Delta-b-\epsilon \sqrt{\frac{4}{\pi} \frac{3}{5}\delta \left (1-\frac{23}{25}\delta\right )} $, i.e.  smaller than the mean energy $ (\Delta-b+\epsilon+m_{\bar{\nu}})$ resulting from antineutrino absorption.\\
 It is amusing to note that, in the presence of the gravitational enhancement, one can achieve a ratio of  rates for neutrino absorption to ordinary electron capture greater than unity, $\lambda_a/\lambda_c\geq1$, for $\Gamma/\epsilon-1\leq10^{-10}$.
\section{Conclusions}
The possibility of observing the cosmic relic neutrinos by their absorption  in a nucleus strongly depends on the mass and the beta-decay properties of the target nuclide  as well as on the  sensitivity and energy precision of the experimental method. The smallest decay energies which have been preferred and considered so far are associated with  tritium and $^{187}$Re in the beta-decay sector, i.e. neutrino absorption, and $^{163}$Ho in the electron-capture sector, i.e. anineutrino absorption. These nuclides, especially tritium and holmium, have been tested as possible candidates for such investigations. A detailed analysis, however, shows that the expected absorption events are very small, considering the feasible amount of a target material, and, for all practical purposes, are not expected to  be observed   in the near future. \\
In this article we have introduced the resonant mechanism of the relic antineutrino absorption. This becomes possible whenever the atomic mass difference is equal to the binding energy of one of the captured electrons, specifically when the excitation energy of the daughter state is equal to the relic antineutrino total energy which eventually is equal to the neutrino rest mass. It was shown that, if the resonant conditions for the antineutrino absorption are met, a considerable enhancement of the associated rates can be obtained, which paves the way towards the experimental observation. The crucial point here is whether such very rigorous energy balance is fulfilled in reality. Obviously, neither tritium and rhenium nor holmium,  mentioned above, are appropriate for the appearance of such a resonance. Hunting for appropriate  candidates led us  consider the case of decay of $^{157}$Tb (71 y) to the excited state of $^{157}$Gd with an energy of 54.5 keV. The energy of the L-transition to this state is not known and presumably ranges from minus 2.5 keV to plus 0.5 keV. Thus such a  resonance may appear.  The strength of the resonance depends on the exact value of this energy around zero. The calculations done for the set of excitations of daughter nuclide $^{157}$Gd  with the excitation energies from $\epsilon$ = 50 to 500 eV (above the 54.5 keV)  shows that the ratio probabilities of relic antineutrino absorption to the ordinary electron L-capture ranges  from $ 10^{-7}$ to $ 10^{-10}$, on the  optimistic assumption  of the gravitational enhancement of the relic neutrinos density by a factor of $10^{6}$. Rough estimations show that the possible number of relic antineutrino capture events could reach a level of hundreds per year for $\epsilon$= 50 eV with a quite reasonable amount of the produced target nuclide ($^{157}$Tb). The terbium nuclides suit for exploration with both the Penning Trap Mass Spectrometry (PT-MS) and the Micro-Calorimetry (MC). The former should provide an answer to the question of how strong the resonance enhancement is expected, whereas the latter should analyze the bolometric spectrum in the tail of the peak corresponding to  L-capture to the excited state in order to observe the relic antineutrino events.   Thus, further considerations of the suitability of $^{157}$Tb for relic antineutrino  detection via the resonant enhancement should be forwarded to the PT-MS and MC teams.

{\bf Acknowledgments}: Both authors are thankful to K. Blaum, S. Eliseev and L. Gastaldo for valuable discussions and to P. Filianin for his assistance in the  preparation of the figures.  The research of one of us (JDV) has been co-financed by the European  Union Social Fund (ESF) and the Greek national funds through the
 program THALIS  of the Hellenic Open University: Development and Applications of Novel Instrumentation and Experimental Methods in Astroparticle  Physics. Yu N would like to thank Max-Planck Institute for Nuclear Physics in Heidelberg for warm hospitality, and Extreme Matter Institute (EMMI) and Russian Minobrnauki (project 2.2) for their support.

  \end{document}